\documentclass[5p,times]{elsarticle}

\usepackage{amsmath}

\begin{document}

\begin{frontmatter}

\title{Operator representation and logistic extension of elementary cellular automata}

\author[marseille]{M. Ibrahimi}
\author[unam,eee]{A. G\" u\c cl\" u}
\author[unam]{N. Jahangirov}
\author[thk]{M. Yaman}
\author[unam,physics]{O. G\"ulseren}
\author[unam,neuroscience]{S. Jahangirov}
\ead{seymur@unam.bilkent.edu.tr}

\address[marseille]{Centre de Physique Th\' eorique (CPT), Turing Center for Living Systems, Aix Marseille Universit\'e, 13009 Marseille, France}
\address[unam]{UNAM-Institute of Materials Science and Nanotechnology, Bilkent University, Ankara 06800, Turkey}
\address[eee]{Department of Electrical and Electronics Engineering, Bilkent University, 06800, Ankara, Turkey}
\address[thk]{Department of Aeronautical Engineering, University of Turkish Aeronautical Association, 06790, Ankara, Turkey}
\address[physics]{Department of Physics, Bilkent University, Ankara 06800, Turkey}
\address[neuroscience]{Interdisciplinary Graduate Program in Neuroscience, Bilkent University, 06800, Ankara, Turkey}

\begin{abstract}
We redefine the transition function of elementary cellular automata (ECA) in terms of discrete operators. The operator representation provides a clear hint about the way systems behave both at the local and the global scale. We show that mirror and complementary symmetric rules are connected to each other via simple operator transformations. It is possible to decouple the representation into two pairs of operators which are used to construct a periodic table of ECA that maps all unique rules in such a way that rules having similar behavior are clustered together. Finally, the operator representation is used to implement a generalized logistic extension to ECA. Here a single tuning parameter scales the pace with which operators iterate the rules. We show that, as this parameter is tuned, many rules of ECA undergo multiple phase transitions between periodic, locally chaotic, chaotic and complex (Class 4) behavior.
\end{abstract}

\end{frontmatter}

Emergence, a semantic gap between behavior and interactions, is a hallmark of dynamical systems. Examples of emergent behavior include: exchanging deals between thousands of agents/companies making up the whole stock market; interactions between alternately expressing genes generating juxtaposed biological forms; series of firings between peculiarly interconnected neurons breeding functional activities in the brain. Given the particularities of such systems (multiple levels of hierarchies, events and processes operating at broadly different space/time scales), complexity science has adopted cellular automata (CA) \cite{neumann} as much simpler computational models to specifically target the semantic gap between individual and global degrees \cite{wolfram1,deutsch,chopard}. These agent based models operate in fully discrete domains and are known to generate large scale types of behavior only through local interactions (rules) \cite{wolfram3,lizard}.

While aiming to acquire a generic understanding applicable to all dynamical systems, a main hypothesis has gathered several attempts to bridge the spatio-temporal patterns observed in CA (phenotype) with their rule space (genotype). The leading apprehension is Wolfram's classification which postulates that the asymptotic behavior of a dynamical system lies in one of these four classes: homogeneous, periodic, aperiodic and complex behavior \cite{wolfram1,wolfram2}. He introduced elementary cellular automata (ECA) as a paradigmatic simple set of rules which comprise all these types.

Several studies have attempted to understand how distinct groups of rules act to generate similar types of asymptotic behavior, eventually making up the four classes. Langton's method of labelling a parameter out of a unit interval to a certain rule is a simple, yet often efficient approach for predicting the class \cite{langton}. Also, further approaches have introduced entropies, mean field descriptions or network analyses that help to understand the relation between patterns and their respective rules \cite{packard,network,gutowitz,binder,politi}. However, these analyses are either too generic to hold for, or too specific to apply to larger families of CA. Consequently, the quest for generalizing a method to any dynamical lattice system remains challenging and this motivates the need to approach CA rules in the light of a different perspective.

Using the ECA set as an example, we suggest a fundamental approach that redefines the transition function based on a simple intuition gained by visual inspection of the system scale dynamics. This approach brings the microscopic information closer to the large scale dynamics and thus helps understand the properties of a system based on its ``first principles".
 
Having distilled the interactions into iterative operations, we employ these operations and the symmetries of the system to rewrite the transition function in a more intuitive notation. Our approach provides a framework that links the similarities and differences observed at the phenotype level to an operator based translation of the rule space. Furthermore, this framework enables us to implement a generalized logistic extension to ECA \cite{jahangirov} where a single parameter scales the pace with which operators iterate the system. As a result, the binary state space of ECA is expanded into a Cantor set and in turn we get a chance to observe transitions between classes \cite{wolfram3}. In particular, we reveal several complex (Class 4) instances that are not reachable in the standard ECA. More interestingly, the behavioral difference between some rules sharing similar genetic code is diminished upon the logistic extension. 

ECA are time dependent one dimensional infinite strings of sites $S^t = \{s_{n} ^t\}|_{n = -\infty} ^\infty$ of a binary state space $s_i \in \{ 0 , 1 \}$. In ECA a Rule defines how the value of a certain site is iterated $s^{t+1} _i = f_S s^{t} _i$ based on its current value and the values of its nearest neighbors, through a transition function $ f_S (s^{t} _{i-1}, s^{t} _i, s^{t} _{i+1}) $. Given the binary state space, there are eight possible configurations of a three site neighborhood, resulting in $2^8 = 256$ possible mappings, i.e rules. Mapping of Rule~30 is shown as an example in Fig.~1(a). The name ``30" of this rule comes from the binary to decimal transformation of the string 00011110 obtained from the particular mapping of the eight configurations listed in the order shown in Fig.~1(a). ECA possesses two important symmetries: complementary and mirror. Here complementary means flipping the state in every site of the array. Hence, if Rule~A and Rule~B are complementary (mirror) symmetric, then running Rule~A with a certain initial string will give the complementary (mirror) image of running Rule~B with the complementary (mirror) version of that string. When these symmetries are taken into account, the number of unique rules reduces to 88 (and not 64 since mirror and/or complementary symmetries of certain rules are equivalent to themselves).

Simple observations on ECA runs reflect visual structures of uniform, stable, oscillatory or irregular patterns. These structures are prone to a mixture of three types of fundamental iterations [$s^{t} _i\rightarrow s^{t+1} _i $], namely: decay [$0|1\rightarrow0$], stability [$0(1)\rightarrow0(1)$], and growth [$0|1\rightarrow1$]. Note that this approach becomes evident in a numerical representation of the state space, and it is the key in translating the rules into discrete operators. We first use the mirror symmetry of these systems to regroup the eight possible configurations into symmetric and asymmetric sets, as seen in Fig.~1(a). Then, within each set, we group complementary configurations together, leading to four groups (denoted by Roman numerals in four different colors). Each group has central cells with values 0 and 1 that are mapped in four different ways: [$0\rightarrow0$,~$1\rightarrow0$], [$0\rightarrow0$,~$1\rightarrow1$], [$0\rightarrow1$,~$1\rightarrow0$], and [$0\rightarrow1$,~$1\rightarrow1$]. We call these double mappings operators, and conveniently name them as Decay (D), Stability (S), Oscillation (O), and Growth (G), respectively. Note that the oscillation operator is a compound of decay and growth iterations. Four groups and four possible operators cover all 256 rules ($4^4 = 2^8$). The operator representation of Rule~30 becomes DSOG. Symmetric counterparts of rules are easily constructed in operator representation. As shown in Fig.~1(b), to get the mirror symmetry of a rule, one needs to switch the operators in the group~III and group~IV. To get the complementary of a rule, one needs to replace all D operations (if any) with G and vice versa. Symmetries of the Rule~30 (DSOG) found by these transformations are presented as an example in Fig.~1(c).

\begin{figure}
\includegraphics[width=8.5cm]{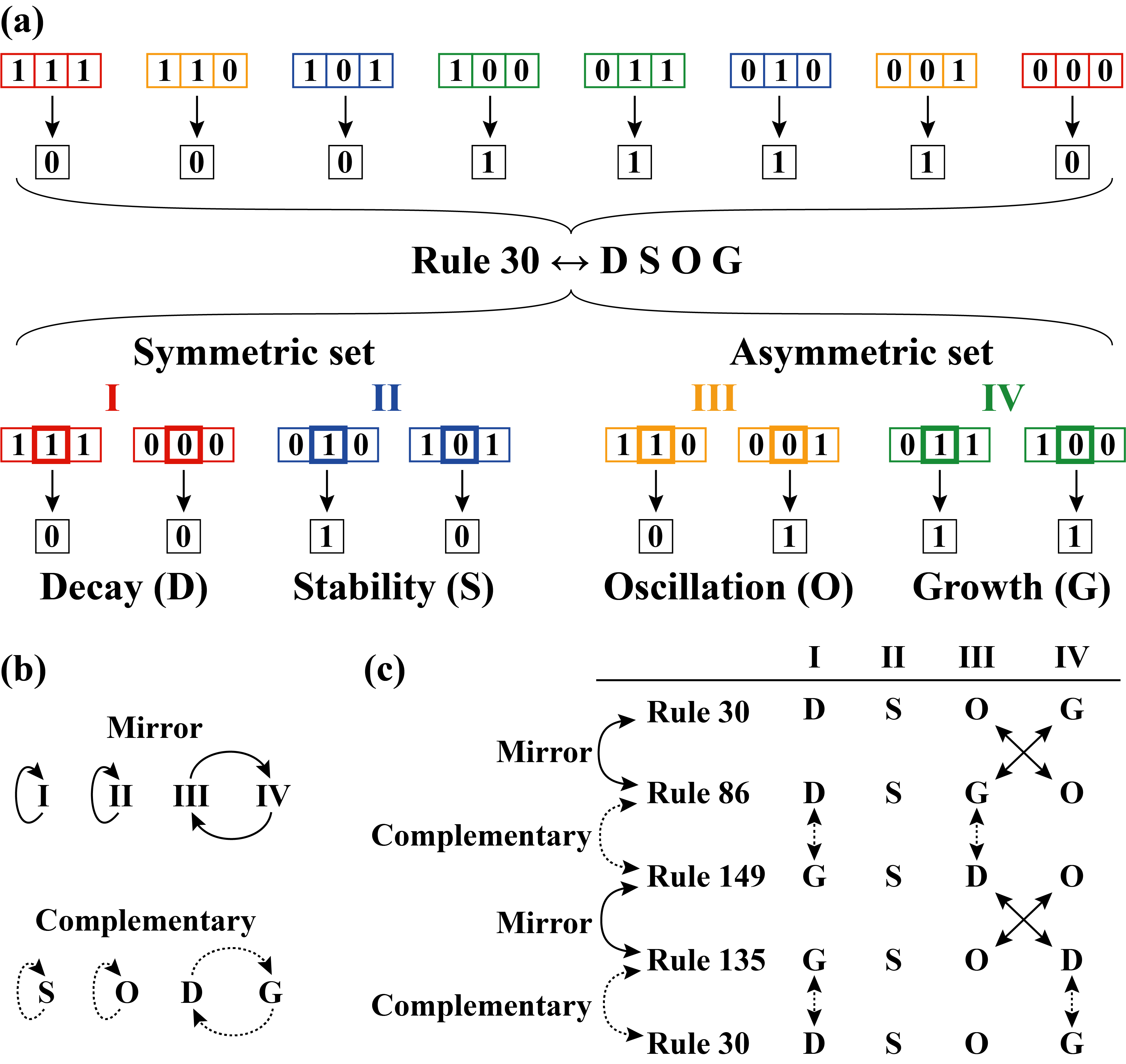}
\caption{(a) Representation of the Rule 30 in terms of operators. (b) Transformations needed to switch between mirror and complementary symmetries of a rule. (c) Switching between the Rule 30 and its symmetries using operator representation.}
\end{figure}
 
Symmetric (I and II) and asymmetric (III and IV) sets of operators are decoupled from each other with respect to both mirror and complementary transformations. Hence, it is instructive to arrange ECA in a ``periodic table'' by placing possible symmetric sets as abscissa and asymmetric sets as ordinate. However, using all 16 pairs of operations in both axes leads to many repetitions of rules that are identical under mirror and complementary transformations. This can be avoided by realizing that, for example, a symmetric set ``DO'' becomes ``GO'' under complementary transformation while remaining the same under mirror transformation. Omitting one of these pairs erases a whole column of repetitions. Continuing in this fashion one can reach at a 10$\times$10 table that has all 88 unique Rules with only 12 repetitions. While constructing this table, one needs to decide which repeating columns to erase and how to arrange the rows and the columns that are left at the end. The table that we have constructed, after evaluating numerous options based on mathematical and aesthetic criteria,  is presented in Fig.~2. The 12 repetitions that appear at the corners of the table are removed for clarity. Note  that, every adjacent row and column share at least one common operator which means that every adjacent rule on the Table share at least three common operators.

\begin{figure*}
\includegraphics[width=17cm]{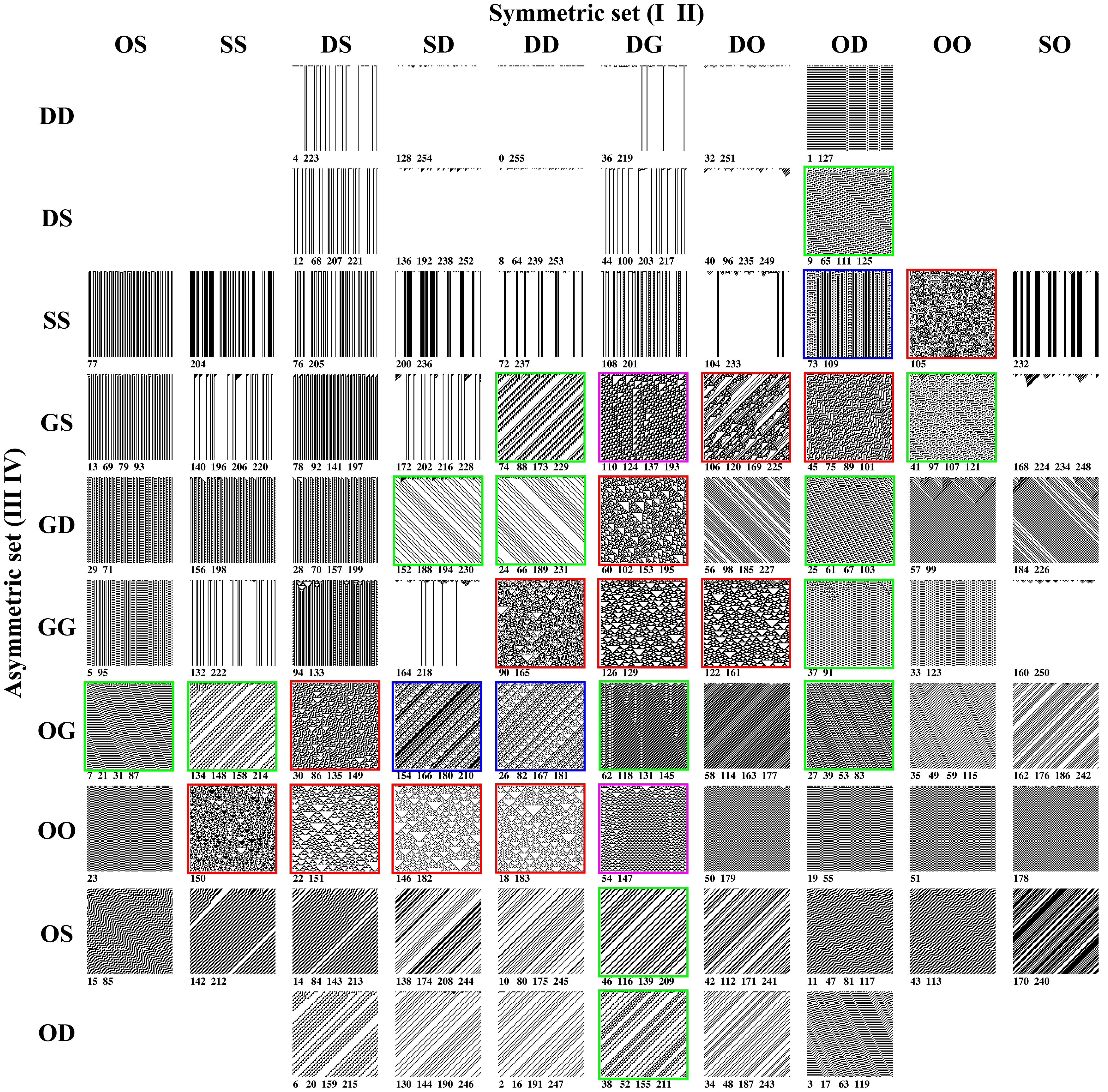}
\caption{Periodic table of the elementary cellular automata (ECA). Rules corresponding to operators representations (in the order I, II, III, IV) and their mirror and complementary counterparts (if different) are presented below each box in increasing order. Each box presents a run starting with a random sequence of 100 binary digits evolved for 100 time steps according to the Rule that is named by the smallest number. Periodic boundary condition is used. Chaotic, locally chaotic and complex rules are highlighted with red, blue and purple squares, respectively. Rules that acquire aperiodic behavior upon the logistic extension are highlighted with green squares.}
\end{figure*}

The Periodic Table presented in Fig.~2 offers a systematic ``bird's eye'' view of all 88 unique rules of ECA. Rules dominated by similar simple patterns (homogeneous, vertical lines, diagonal lines, horizontal stripes) tend to appear together. The rules that show rich behavior populate the ``fertile crescent'' along the diagonal where simple rules with contradicting patterns are expected to overlap. Among these rich rules, the ones that have common features are also brought together. Rule pairs 18, 146 and 122, 126 are striking examples of this. Despite the chaotic nature of these rules, starting a run with one of them and switching to the other rule results in the same pattern that is produced without the switching. This is because, Rule~18 (122) and Rule~146 (126) share the same mapping, except for the configuration 111 (010) which is mapped to 0 in the former and 1 in the latter. This 111 (010) configuration is ``washed out'' in a few steps and is never visited again. This effect is also present if one starts with the Rule~26 and continues with the Rule~154 but not the other way around.

The Periodic Table of ECA also resonates with the findings of Li and Packard \cite{packard} in their classic study on the structure of the ECA rule space. They found two clusters of chaotic rules (in this context it includes the complex rules 54 and 110). Chaotic A includes Rules~18, 22, 30, 54, 146, and 150 while Chaotic B has Rules~60, 90, 106, 110, 122, and 126. As seen in Fig~2, they appear as clusters at the bottom left and top right of the ``fertile crescent'', respectively. The authors found Rule~45 to be separated from the clusters but in our Table we find it connected to the cluster B. Furthermore, clusters A and B are connected over a bridge of locally chaotic Rule~26 in the Table. There are no other chaotic rules in the row and the column of the  Rule~105, which was also found to be isolated by Li and Packard, but it is connected to the cluster B over a bridge of locally chaotic Rule~73.

The operator representation can further illuminate the studies on the computational irreducibility of ECA. In particular, it is interesting to examine the rules that are detached from the coarse-graining network investigated by Israeli and Goldenfeld \cite{israeli}. They have shown that, Rule~105 can be course grained by the Rule~150. In the operator representation, these rules appear as OOSS and SSOO, respectively. Furthermore, both DGDG (Rule 60) and DDGG (Rule 90) can be coarse-grained by themselves. Finally, the authors were unable to coarse-grain four unique rules: 30, 45, 106 and 154. In the operator representation, they happen to be DSOG, OGDS, OSGD, and SDOG. These make up four unique rules that involve all four operators while avoiding two complementary symmetric operators (D and G) in the same mirror symmetric set. In other words, the rules that were found to be irreducible are the ones that appear the most asymmetric in the operator representation. We believe that these mere observations can guide further studies in this subject.

\begin{figure}
\includegraphics[width=8.5cm]{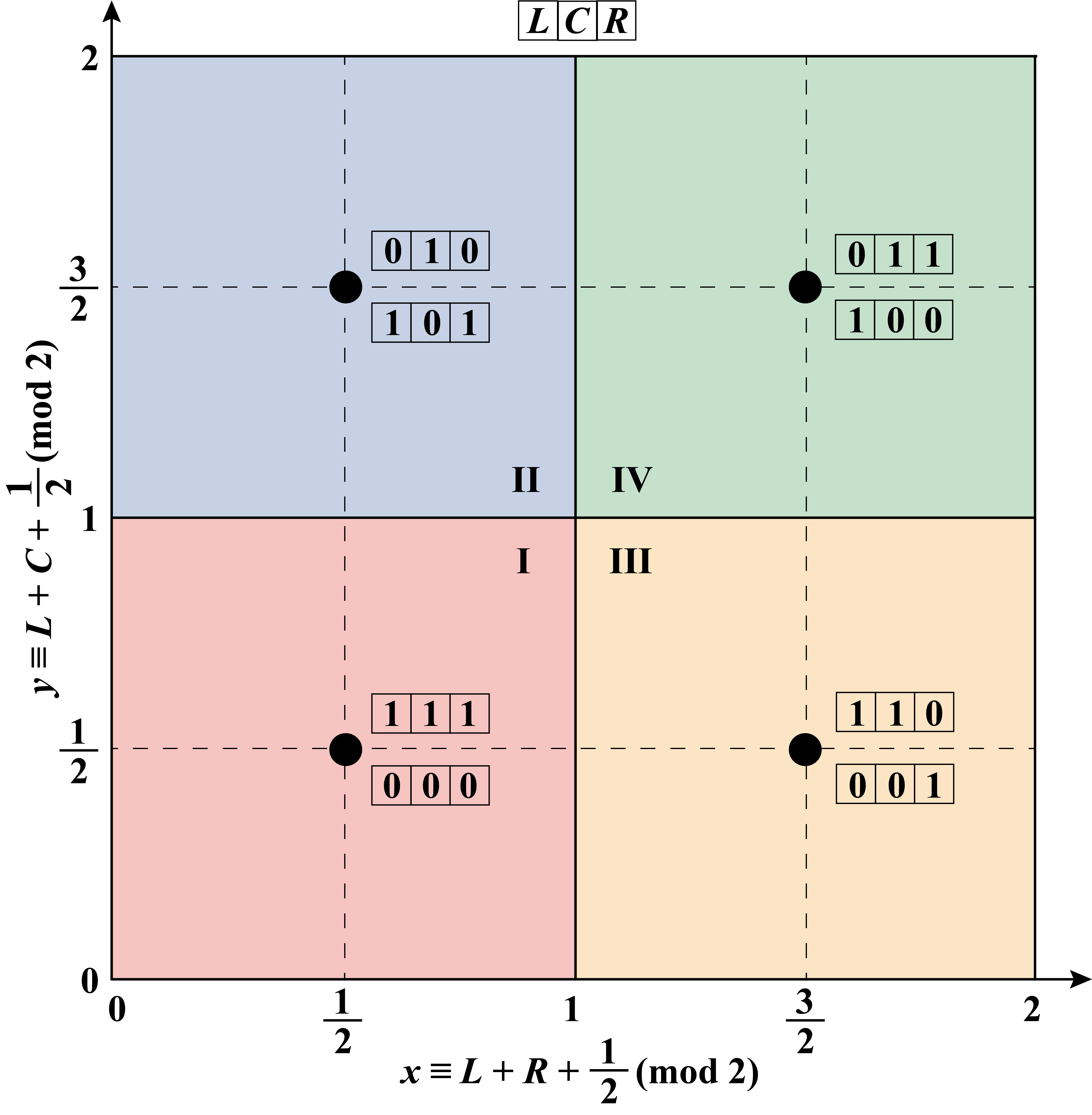}
\caption{Definition of the operation regions based on a configuration. $L$, $C$ and $R$ correspond to the values at the left, center and right cell of a configuration.}
\end{figure}

Recently, we have introduced the logistic extension of two outer-totalistic CA: Game of Life and Rule~90. This extension is achieved via introduction of a parameter, $\lambda$, that tunes the dynamics of CA. $\lambda = 1$ corresponds to the original binary version of the studied systems. As $\lambda$ is tuned below 1, the binary state space extends into a Cantor set and the systems expand their complexity through series of deterministic transitions \cite{jahangirov}. In particular, the Rule 90 which is aperiodic at $\lambda = 1$ shows complex (or Class 4) behavior at $\lambda \sim 0.6$. The operator representation presented here enables us to go beyond the outer-totalistic rules and generalize the logistic extension to all ECA. We first define four regions of operation for each group (I, II, III and IV) as shown in Fig.~3. The coordinates of a configuration $[L, C, R]$ (denoting left, center and right sites, respectively) defined as the sums  $x \equiv L+R+\frac{1}{2} \pmod{2}$ and $y \equiv L+C+\frac{1}{2} \pmod{2}$ determine in which operation region it falls. As shown in Fig.~3, the eight possible binary configurations appear at the centers of the regions that correspond to their group definitions shown in Fig~1(a). Hence, the configuration $[L, C, R]$ determines the operation region which in turn determines the corresponding operator based on the rule at hand. Depending on the operator, the value of a site is updated according to the following formulae:

\begin{equation*}
\begin{split}
\text{Decay} ~ \Rightarrow ~ & s^{t+1} = (1 - \lambda) s^t \\
\text{Stability} ~ \Rightarrow ~ & s^{t+1} = s^t \\
\text{Oscillation} ~ \Rightarrow ~ & s^{t+1} =
    \begin{cases}
      (1 - \lambda) s^t + \lambda, & \text{if}\ s^t \leq \frac{1}{2} \\
      \\
      (1 - \lambda) s^t, & \text{if}\ s^t > \frac{1}{2}
    \end{cases}  \\
\text{Growth} ~ \Rightarrow ~ & s^{t+1} = (1 - \lambda) s^t + \lambda \\
\end{split}
\end{equation*}

where $s^t$ and $s^{t+1}$ are the values of the central site at the current and the next time step, respectively. These equations, consistent with the operator notation, make up the new form of the transition function. Note that this generalization is consistent with the special case of the Rule~90 that we have reported earlier \cite{jahangirov}.

Significant changes in dynamics can occur when $x$ or $y$ passes over from one region to another. This happens when the sum $L+C$ or $L+R$ is equal to the critical thresholds 0.5 or 1.5. The values that $L$, $C$, and $R$ can take is dictated by the $\lambda$-dependent Cantor set. Hence, one can expect these changes at the values of $\lambda$ that mark the equality of binary sums to the critical thresholds. As $\lambda$ is tuned below 1, the first time such a transition occurs is when $2 \lambda = 1.5$. After this point, some of the rules start behaving differently than their original version. 

Rules that exhibit chaotic, locally chaotic or complex behavior pass through multiple phase transitions while going between these regimes. As seen in Fig.~4, chaotic Rule 18 becomes complex at $\lambda = 0.73$ mimicking (but not exactly copying) the complex patterns seen in one of its neighbors, Rule 54. Another locally chaotic rule close by, Rule 82, also mimics the Rule 54 behavior at $\lambda = 0.74$.

Logistic extension breaks the symmetry between mirror rules because of the left-right asymmetry in the sum $L+C$. This is clear in the distinct behavior of Rule~26 (the mirror symmetry of Rule~82) which has a mixture of chaotic and locally chaotic behavior at $\lambda = 0.74$. However, complementary rules behave in the same way under the logistic extension. For example, the behavior of complementary Rules~90 and 165 are the same at $\lambda = 0.6$ \cite{jahangirov}. Rules that originally have complex behavior may remain complex while having noticeable changes in their dynamics, for example Rule 54 at $\lambda = 0.74$. They also can become locally chaotic like Rule 110 at $\lambda = 0.74$ or become chaotic like Rule~124 (not shown in Fig.~4) which resembles its neighboring chaotic Rule~60 at $\lambda = 0.72$. Rules that are chaotic or locally chaotic can behave in a complex fashion as exemplified by Rule~86 (mirror symmetry of the Rule~30) at $\lambda = 0.68$ and Rule~154 at $\lambda = 0.68$, respectively.

Some of the Rules that are originally periodic can acquire aperiodic behavior. For example, Rule~38 becomes locally chaotic at $\lambda \sim 0.69$ and chaotic at $\lambda \sim 0.61$. Periodic rules can also become complex, for example Rule~37 and Rule~46 at $\lambda = 0.72$, as shown in  Fig.~4. Rules that gain aperiodic behavior upon the logistic extension are highlighted by the green squares in Fig.~2. Note that, these rules are adjacent to the rules that are originally aperiodic.

\begin{figure}
\includegraphics[width=8.5cm]{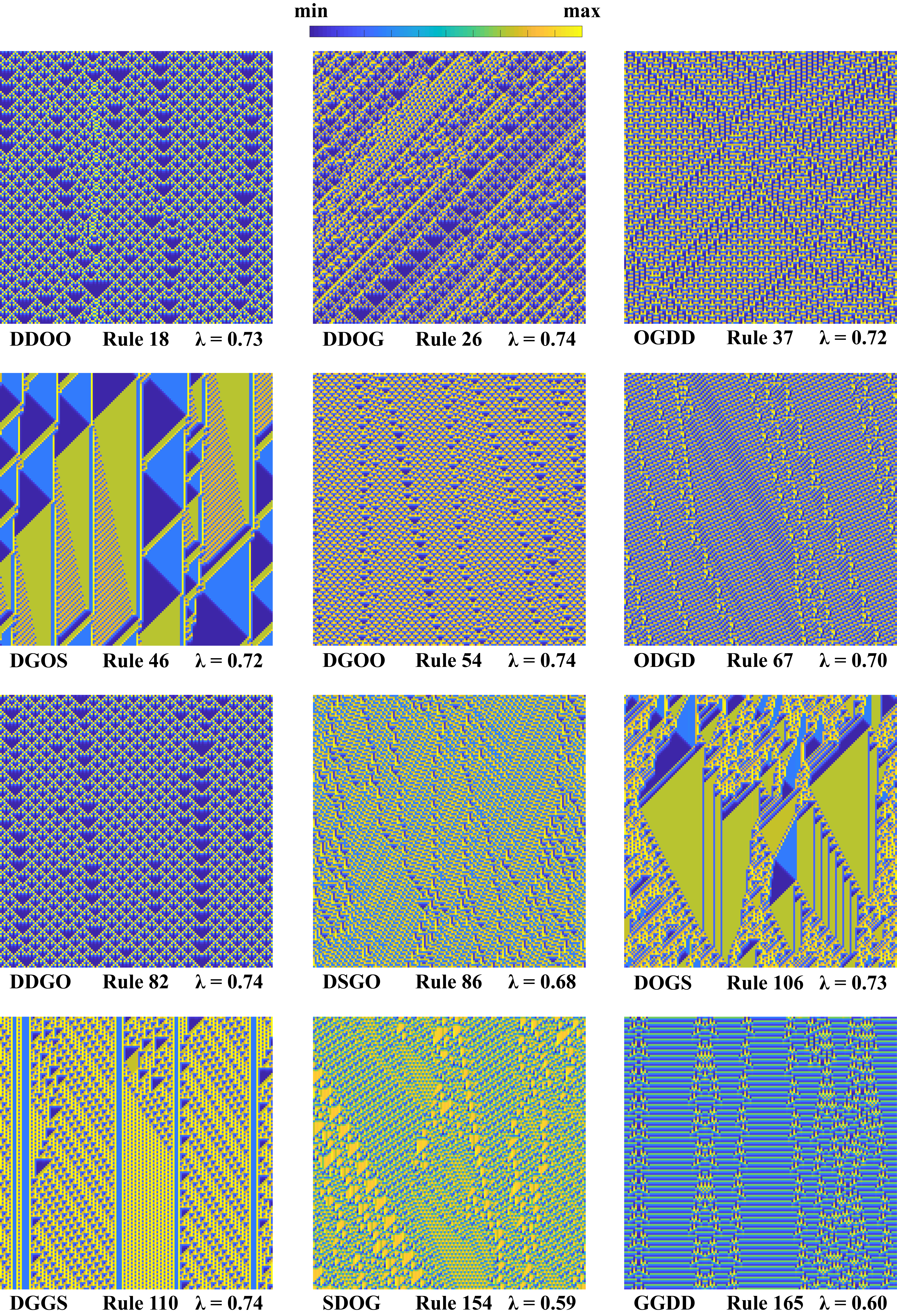}
\caption{150$\times$150 cells snapshots at a later stage of a 1000$\times$1000 simulation for various Rules with given values of $\lambda$. The color bar shown at the top maps the range between the minimum and maximum cell values for each snapshot. Both conventional and operator representation of the rules are given below each panel.}
\end{figure}

In summary, to understand the disjunctive and connective (diverse and unifying) nature of ECA rules, we redefine the transition function by introducing an operator-based notation. This allows one to organize the rules in a periodic table where underlying connections between their macroscopic behaviors and genetic codes emerge. Furthermore, we introduce a tuning parameter which controls the rate of rule iterations. This parameter extends the range of behavior that ECA can offer while generating inter-class transitions and disclosing inert behaviors of periodic rules. We believe that logistic extension to the operator-based representation may be useful to explore hidden features in other complex systems, such as discrete lattice models \cite{kaneko} and boolean genetic networks \cite{kauffman}.

S. J. acknowledges support from the Turkish Academy of Sciences - Outstanding Young Scientists Award Program (T\"UBA-GEB\. IP).

\end{document}